%% file: main.tex
\documentclass[]{aastex631}

\shorttitle{An Ocean Expedition to Retrieve Fragments of CNEOS 2014-01-08}
\shortauthors{Siraj, Loeb, \& Gallaudet}
\graphicspath{{./}{figures/}}

\begin{document}

\title{An Ocean Expedition by the Galileo Project to Retrieve Fragments of the \\ First Large Interstellar Meteor CNEOS 2014-01-08}

\email{aloeb@cfa.harvard.edu, amir.siraj@cfa.harvard.edu}

\author{Amir Siraj}
\affiliation{Department of Astronomy, Harvard University, 60 Garden Street, Cambridge, MA 02138, USA}

\author{Abraham Loeb}
\affiliation{Department of Astronomy, Harvard University, 60 Garden Street, Cambridge, MA 02138, USA}

\author{Tim Gallaudet}
\affiliation{Ocean STL Consulting, LLC}





\input{tex/abstract}



\input{tex/intro}

\input{tex/properties}
\input{tex/expedition}
\bibliographystyle{aasjournal}
\bibliography{bib.bib}

\end{document}

%% file: tex/abstract.tex
\begin{abstract}
The earliest confirmed interstellar object, `Oumuamua, was discovered in the Solar System by Pan-STARRS in 2017, allowing for a calibration of the abundance of interstellar objects of its size $\sim 100\;$ m. This was followed by the discovery of Borisov, which allowed for a similar calibration of its size $\sim 0.4 - 1 \mathrm{\; km}$. One would expect a much higher abundance of significantly smaller interstellar objects, with some of them colliding with Earth frequently enough to be noticeable. Based on the CNEOS catalog of bolide events, we identified in 2019 the meteor detected at 2014-01-08 17:05:34 UTC as originating from an unbound hyperbolic orbit with 99.999\% confidence. In 2022, the U.S. Department of Defense has since verified that ``the velocity estimate reported to NASA is sufficiently accurate to indicate an interstellar trajectory,'' making the object the first detected interstellar object and the first detected interstellar meteor. Here, we discuss the dynamical and compositional properties of CNEOS 2014-01-08, and describe our plan for an expedition to retrieve meteoritic fragments from the ocean floor. CNEOS 2014-01-08 is an outlier both in terms of its LSR speed (shared by less than 5\% of all stars) and its composition (tougher than all 272 other bolides in the CNEOS database). Our plan is to mobilize a ship with a magnetic sled off the coast of Papua New Guinea to collect fragments of CNEOS 2014-01-08.

\end{abstract}

%% file: tex/intro.tex
\section{Introduction}

Two interstellar objects have been detected so far in the solar system through their reflection of sunlight. `Oumuamua was the first interstellar object detected in the Solar System by Pan-STARRS \citep{2017Natur.552..378M, 2018Natur.559..223M}. Several follow-up studies of `Oumuamua were conducted to better understand its origin and composition \citep{2017ApJ...851L..38B, 2017RNAAS...1...13G, 2017ApJ...850L..36J, 2017RNAAS...1...21M, 2017ApJ...851L...5Y, 2018ApJ...852L...2B, 2018NatAs...2..133F, 2018AJ....156..261T, 2018ApJ...868L...1B, 2018ApJ...860...42H, 2019ApJ...872L..10S, 2019RNAAS...3...15S, 2019ApJ...876L..26S}. Its size was estimated to be 20m -- 200m, based on Spitzer Space Telescope constraints on its infrared emission given its temperature \citep{2018AJ....156..261T}. The discovery of `Oumuamua was followed by that of the second interstellar object, Borisov, in 2019 \citep{2020NatAs...4...53G}. The size of Borisov's nucleus was estimated to be $0.4 - 1 \mathrm{\; km}$ \citep{2020ApJ...888L..23J, 2021MNRAS.507L..16S}. 

One interstellar object, the earliest detected, was observed as it burned up in the Earth's atmosphere. CNEOS 2014-01-08, originating from outside the soalr system, was $\sim 0.5 \mathrm{\; m}$ in size and burned up in the Earth's atmosphere off of the coast of Papua New Guinea in 2014 \citep{2019arXiv190407224S}. This discovery was confirmed on April 6, 2022, when the United States Department of Defense (US DoD) released a formal letter validating the interstellar origin of CNEOS 2014-01-08 at the 99.999\% confidence level \citep{shaw_2022}.\footnote{\url{https://lweb.cfa.harvard.edu/~loeb/DoD.pdf}}

The discovery of an interstellar meteor heralds a new research frontier, in which the Earth serves as a fishing net for massive interstellar objects.  As a result of encountering Earth and rubbing against its atmosphere, an interstellar object burns up in a bright fireball. This fireball is detectable by satellites or ground-based sensors even for relatively small interstellar objects like CNEOS-2014-01-08, which created a fireball carrying a few percent of the energy of the Hiroshima bomb. This size scale is a two orders of magnitude smaller than `Oumuamua, which was discovered by the Pan-STARRS telescope through its reflection of sunlight. This alternative detection method allows existing survey telescopes to discover only objects larger than a football field, within the orbit of the Earth around the Sun. 

The interstellar meteor discovery is very important from another perspective. A space mission to visit on an interstellar object like `Oumuamua and return a sample of it to Earth, similar to the OSIRIS-REx\footnote{\url{https://www.asteroidmission.org/}} mission that landed on the asteroid Bennu and will return material from it in September 2023, would cost $\sim \$1 \mathrm{B}$. But at a cost that is $\sim 10^4$ times smaller, one could retrieve fragments left over from an interstellar meteor and study them in laboratories on Earth. Here, we explore the properties of the first interstellar meteor, CNEOS-2014-01-08, and discuss the possibility of recovering debris from it on the ocean floor.

%% file: tex/properties.tex
\section{Properties of the First Interstellar Meteor}

\subsection{Dynamics}

Recently, the U.S. Department of Defense, which houses the classified data pertaining to the uncertainties involved in the CNEOS 2014-01-08 detection, released a public statement dated March 1, 2022, and addressed to the NASA Science Mission Directorate, referencing this discovery preprint and mentioning our analysis that the meteor originated from an unbound hyperbolic orbit with 99.999\% confidence \citep{shaw_2022}. The letter then states: ``Dr. Joel Mozer, the Chief Scientist of Space Operations Command, reviewed analysis of additional data available to the Department of Defense related to this finding. Dr. Mozer confirmed that the velocity estimate reported to NASA is sufficiently accurate to indicate an interstellar trajectory'' \citep{shaw_2022}. The statement thereby confirms interstellar origin of the 2014-01-08 meteor, and directly references the discovery by Amir Siraj \& Abraham Loeb \citep{2019arXiv190407224S}.

\begin{figure}
  \centering
  \includegraphics[width=.9\linewidth]{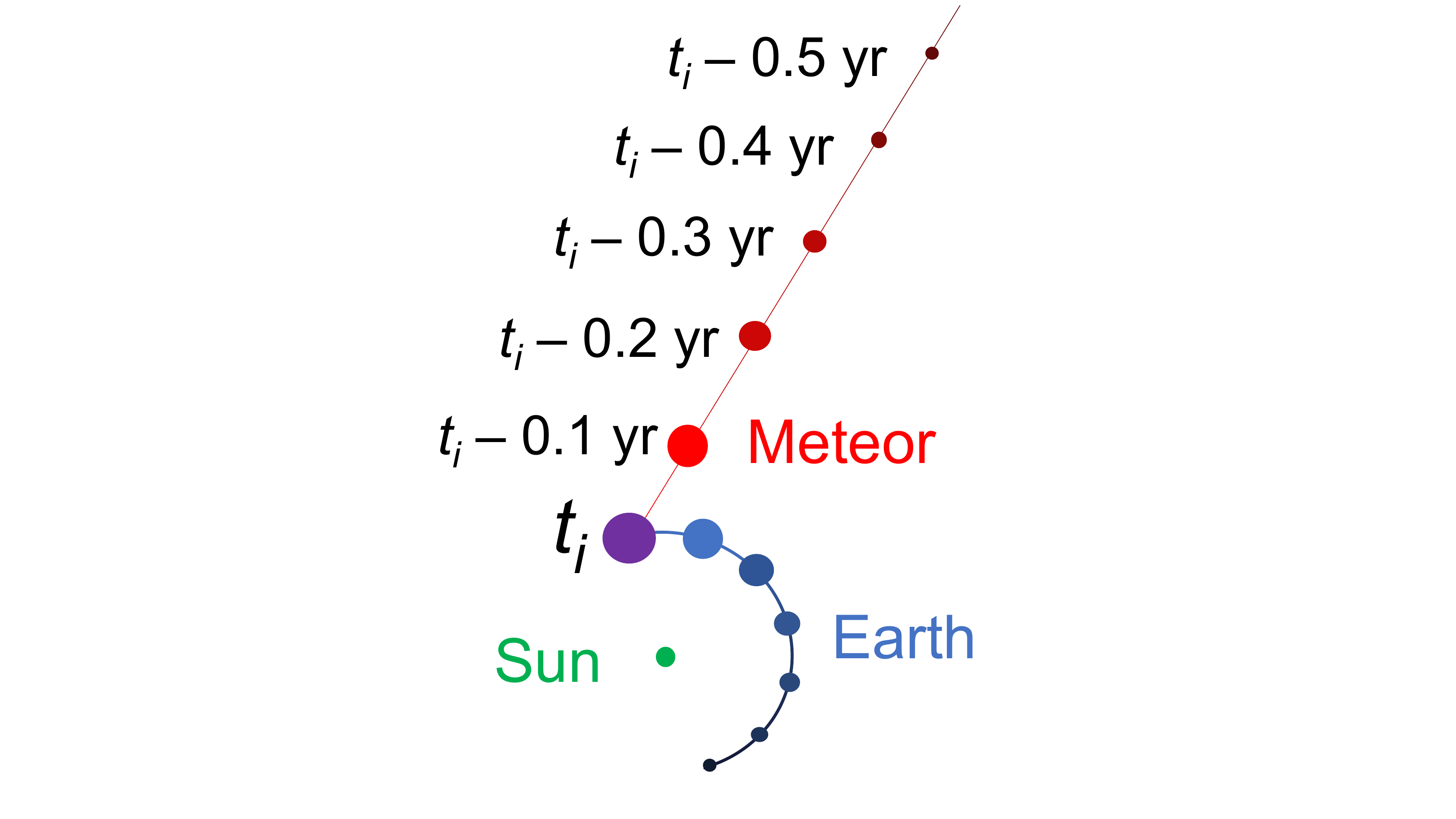}
    \caption{Trajectory of the January 8, 2014 meteor (red), shown intersecting with that of Earth (blue) at the time of impact, $t_i =$ 2014-01-08 17:05:34.}
    \label{fig:trajectory}
\end{figure}

The heliocentric orbital elements of the meteor at time of impact are as follows: semi-major axis, $a = -0.47 \pm 0.15 \;$ AU, eccentricity, $e = 2.4 \pm 0.3$, inclination $i = 10\pm2^{\circ}$, longitude of the ascending node, $\Omega = 108\pm1^{\circ}$, argument of periapsis, $\omega = 58\pm2^{\circ}$, and true anomaly, $f = -58\pm2^{\circ}$. The trajectory is shown in Fig.~\ref{fig:trajectory}. The origin is towards R.A. 
$49.4 \pm 4.1^{\circ}$ and declination $11.2 \pm 1.8^{\circ}$. The heliocentric incoming velocity at infinity of the meteor in right-handed Galactic coordinates is $v_{\infty}\mathrm{(U, V, W) = (32.7 \pm 5.8, -4.5 \pm 1.5, 26.1 \pm 2.0)\;}$ $\mathrm{km\;s^{-1}}$, which is $58\pm6\; \mathrm{km\;s^{-1}}$ away from the velocity of the Local Standard of Rest (LSR), $\mathrm{(U, V, W)_{LSR}}$ $= (-11.1, -12.2, -7.3)\;$ $\mathrm{km\;s^{-1}}$ \citep{2010MNRAS.403.1829S}.

Given the impact speed of the meteor, $\sim 44.8 \; \mathrm{km\;s^{-1}}$, and the total impact energy, $4.6 \times 10^{18}$ ergs, the meteor mass was approximately $4.6 \times 10^5 $ g. Slowdown by friction on air implies a higher initial speed above the atmosphere. 

The CNEOS catalog includes bolide events at a relatively high frequency for the past decade, so we approximate the yearly detection rate of interstellar meteors to be at least $\sim 0.1 \mathrm{\: yr^{-1}}$. We estimate the number density of similarly sized interstellar objects by dividing the yearly detection rate by the product of the impact speed of the meteor and the cross sectional area of the Earth, finding the approximate number density of interstellar objects with a similar size and speed relative to the LSR to be, 

\begin{equation}
n\sim\frac{0.1 \: \mathrm{yr^{-1}}}{(13 \: \mathrm{ AU/yr})(5.7 \times 10^{-9} \: \mathrm{AU^2})} \sim 10^6 \: \mathrm{ AU^{-3}}.
\end{equation}

Given 95\% Poisson uncertainties \citep{1986ApJ...303..336G}, the inferred\footnote{Gravitational focusing by the Earth is negligible since the meteor speed exceeds considerably the escape speed from the Earth. The density enhancement due to gravitational focusing by the Sun is well below the uncertainty in the estimated value of $n$, so that our inferred range of local values also corresponds to the density outside of the Solar System.} local number density for interstellar objects of this size is $n = 10^{6{^{+0.75}_{-1.5}}} \; \mathrm{AU^{-3}}$. This figure necessitates $6 \times 10^{22{^{+0.75}_{-1.5}}}$ similarly size objects, or 0.2 -- 20 Earth masses of material, to be ejected per local star. This is at tension with the fact that a minimum-mass solar nebula is expected to have about an Earth mass of total planetesimal material interior to the radius where the orbital speed is $\sim 60 \; \mathrm{km\;s^{-1}}$ \citep{2007ApJ...671..878D}, with similar values for other planetary systems \citep{2004ApJ...612.1147K}. Our inferred abundance for interstellar meteors should be viewed as a lower limit since the CNEOS data might have a bias against detection of faster meteors \citep{2016Icar..266...96B}.

Its $\sim 58\pm6\; \mathrm{km \; s^{-1}}$ deviation from the LSR suggests that it perhaps originated in the thick disk, which has velocity dispersion components of $\mathrm{(\sigma_U, \sigma_V, \sigma_W) = (50, 50, 50)\;} \mathrm{km \; s^{-1}}$  relative to the LSR \citep{2016ARA&A..54..529B}, and slowdown by friction on air implies that the speed relative to the LSR was even higher. However, the ratio of local thick disk stars to thin disk stars is 0.04, making this a minority population. Alternatively, for a parent planetary system with a more typical velocity relative to the LSR, the object could have originated in the deep interior, where the orbital speeds of objects are of the necessary magnitude. Either way, the meteor had an unusual origin.

\subsection{Composition}

Along with the confirmation letter, the US DoD released the unusual light curve for CNEOS 2014-01-08, with three flashes separated from each other by about a tenth of a second. The measured direction of motion for CNEOS 2014-01-08 can be used to calculate the altitude of the three flashes in the explosion and the ambient density of air where they occurred.

When a supersonic meteor moves through air, it is subject to a friction force on its frontal surface area. The force per unit area equals the ambient mass-density of air times the square of the object’s speed. This ram pressure reflects the flux of momentum per unit area per unit time delivered to the object in slowing down its motion. The meteor disintegrates if the ram pressure exceeds the yield strength of the material it is made of, representing the maximum stress that can be applied to it before it begins to deform. The heat released by the friction with air melts the fragments and generates the flashes of light in the fireball. We complete this calculation in \cite{2022RNAAS...6...81S}, as described below. 

First, we convert the optical power reported in the light curve to total power. Using Equation (1) from \cite{2002Natur.420..294B}, combined with the total optical energy for CNEOS 2014-01-08 of $3.1 \times 10^{17} \; \mathrm{erg}$ \citep{2019arXiv190407224S}, we find that the optical efficiency is $\sim 6.9\%$. The total power as a function of time is therefore simply the optical power as a function of time, divided by $6.9\%$.

\begin{figure}
  \centering
  \includegraphics[width=\linewidth]{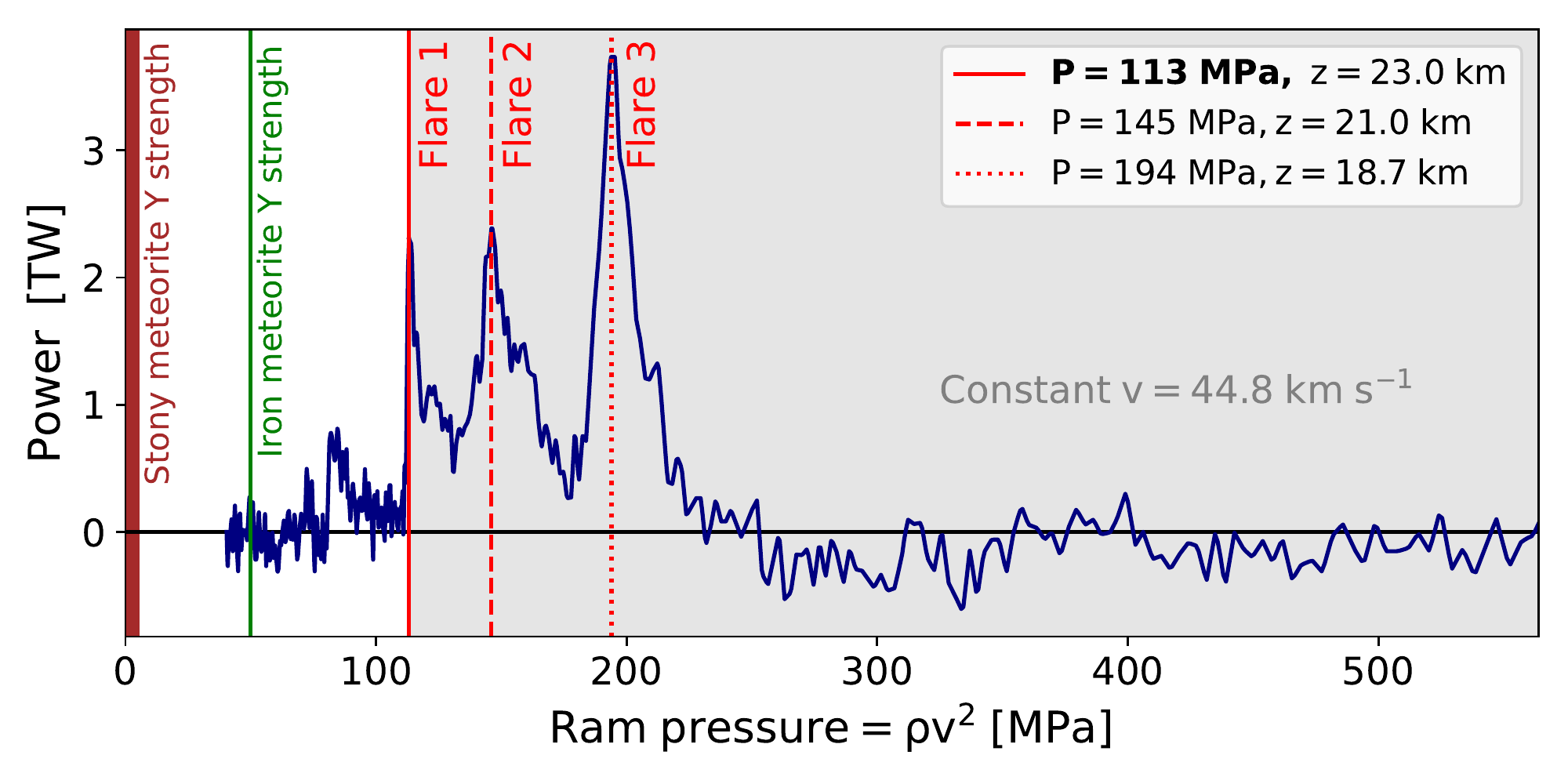}
    \caption{Total power released in the CNEOS 2014-01-08 fireball as a function of ram pressure, $\rho v^2$. Typical stony and iron meteorite yield strengths, $1 - 5 \mathrm{\; MPa}$ and $50 \mathrm{\; MPa}$ respectively, are indicated for convenience of comparison, and the three major flare events are labelled according to the order in which they occurred. We conservatively adopt a constant velocity between the flares. Note that 1 TW = $10^{19} \; \mathrm{erg \; s^{-1}}$ and 1 MPa = $10^7 \mathrm{\; dyne \; cm^{-2}}$.}
    \label{fig:p}
\end{figure}

Next, we derive the ram pressures, $\rho v^2$, corresponding to the three major explosion flares visible in the light curve, where $\rho$ is the ambient mass-density of air and $v$ is the meteor speed. We adopt a straight-line trajectory for CNEOS 2014-01-08 as it moves through the atmosphere at an angle of $\theta = 26.8^{\circ}$ relative to the ground \citep{2019RNAAS...3...68Z}. The velocity and altitude measurements reported by CNEOS, $v_{CNEOS} = 44.8 \mathrm{\; km \; s^{-1}}$ and $z_{CNEOS} = 18.7 \mathrm{\; km}$, correspond to peak brightness,\footnote{\url{https://cneos.jpl.nasa.gov/fireballs/intro.html}} or Flare 3 in Figure \ref{fig:p}. We conservatively adopt a constant speed of $v_{CNEOS} = 44.8 \mathrm{\; km \; s^{-1}}$ between the flares; deceleration due to object breakup between Flare 1 and Flare 3 would lead to greater ram pressures. The $\Delta t_{2,3} = 0.112 \; \mathrm{s}$ delay between Flares 2 and 3, and the $\Delta t_{1,3} = 0.213 \; \mathrm{s}$ delay between Flares 1 and 3, imply that Flares 2 and 3 occurred at $(v_{CNEOS} \Delta t_{2,3} \sin{\theta}) = 2.3 \mathrm{\; km}$ and $(v_{CNEOS} \Delta t_{1,3} \sin{\theta}) = 4.3 \mathrm{\; km}$ above the altitude at which Flare 3 occurred, respectively. This indicates that Flare 1 and Flare 2 occurred at altitudes of $z = 23.0 \mathrm{\; km}$ and $z = 21.0 \mathrm{\; km}$, respectively. 

We adopt the atmospheric density profile of $\rho(z) = \rho_0 \exp{(-z / H)}$, where $\rho_0 = 10^{-3} \mathrm{\; g \; cm^{-3}}$ is the sea-level atmospheric density and $H = 8 \mathrm{\; km}$ is the scale height of the Earth's atmosphere \citep{2005M&PS...40..817C}. The atmospheric densities at which Flare 1, Flare 2, and Flare 3 transpired are $\rho_1 = \rho(23.0 \mathrm{\; km}) = 5.64 \times 10^{-5} \; \mathrm{g \; cm^{-3}}$, $\rho_2 = \rho(21.0 \mathrm{\; km}) = 7.24 \times 10^{-5} \; \mathrm{g \; cm^{-3}}$, and $\rho_3 = \rho(18.7 \mathrm{\; km}) = 9.66 \times 10^{-5} \; \mathrm{g \; cm^{-3}}$. The resulting ram pressures for the three flares are $(\rho_1 v_{CNEOS}^2) = 113 \mathrm{\; MPa}$, $(\rho_2 v_{CNEOS}^2) = 145 \mathrm{\; MPa}$, and $(\rho_3 v_{CNEOS}^2) = 194 \mathrm{\; MPa}$, respectively. Figure \ref{fig:p} displays the flares in terms of total power as a function ram pressure.

Breakup occurs when the yield strength of the impactor $Y_i$ is equivalent to the ram pressure: $Y_i = \rho v^2$ \citep{2005M&PS...40..817C}. CNEOS 2014-01-08's active phase is bracketed by a narrow range of ram pressures, $113-194$ MPa, which translates directly into the constraint on the original object's yield strength. Most conservatively, the yield strength of CNEOS 2014-01-08 was comparable to the ram pressure of Flare 1, $Y_i = (\rho_1 v_{CNEOS}^2) = 113 \mathrm{\; MPa}$.

Based on estimates for comets, carbonaceous, stony, and iron meteorites \citep{1993Natur.361...40C, 1993Natur.365..733S, 1995Icar..116..131S, 2001JMatS..36.1579P}, \cite{2005M&PS...40..817C} established an empirical strength-density relation for impactor density $\rho_i$ in the range $1 - 8 \mathrm{\; g \; cm^{-3}}$. The upper end of this range gives a yield strength of $Y_i \sim 50 \mathrm{\; MPa}$, corresponding to the strongest known class of meteorites, iron \citep{2001JMatS..36.1579P}. Iron meteorites are rare in the solar system, making up only $\sim 5\%$ of modern falls \citep{2006mess.book..869Z}. The CNEOS 2014-01-08 inferred yield strength of at least $Y_i = 113 \mathrm{\; MPa}$ exceeds the typical yield strength of iron meteorites by a factor of $\sim 2$. The observed yield strength is also inconsistent with stony meteorites, which exhibit a range of lower yield strengths by $1 - 2$ orders of magnitude \citep{2001JMatS..36.1579P, 2011M&PS...46.1525P}. Finally, the natural possibilities considered for `Oumuamua's composition are ruled out for CNEOS 2014-01-08 on the basis of insufficient strength, namely a nitrogen iceberg \citep{2021JGRE..12606706J, 2021JGRE..12606807D}, an $\mathrm{H_2}$ iceberg \citep{2020ApJ...896L...8S}, or a fluffy dust cloud \citep{2019ApJ...872L..32M, 2020ApJ...900L..22L}.

Additionally, CNEOS 2014-01-08 experienced slowdown between its atmospheric entry and detonation. We define a slowdown factor $f_s$, derived from Equation (8) in \cite{2005M&PS...40..817C},
\begin{equation}
    f_s(z) = \exp{\left(-\frac{3 \rho(z) C_D H}{4 \rho_i L_0 \sin \theta}\right)},
\end{equation}
where $C_D = 2$ is the drag coefficient and $L_0 = 2 \times (3 E / 2 \pi v_{CNEOS}^2 \rho_i)^{1/3}$ is the diameter of the object, where $E = (3.1 \times 10^{17}/ 6.9\%) \mathrm{\; erg}$ is the total explosion energy. The meteor's speed at an altitude $z$ above the range of breakup altitudes is then $v(z) \sim [f_s(z) v_{CNEOS} / f_s(z_{CNEOS})]$. Adopting a fiducial density of $\rho_i = 8 \mathrm{\; g \; cm^{-3}}$, corresponding to an iron composition and $L_0 \sim 0.5~{\rm m}$, this implies that the impactor's speed at the top of the atmosphere was at least $v(z \rightarrow \infty) = 66.5 \mathrm{\; km \; s^{-1}}$, which is $22 \mathrm{\; km \; s^{-1}}$ or $48 \%$ faster than $v_{CNEOS} = 44.8 \mathrm{\; km \; s^{-1}}$, the impact speed used to evaluate the orbit and determine the interstellar origin of CNEOS 2014-01-08 \citep{2019arXiv190407224S}. A lower value of $\rho_i$ would lead to greater slowdown. This increase in geocentric impact speed makes the interstellar origin of CNEOS 2014-01-08 clearer, and its motion relative to the local standard of rest even more anomalous \citep{2019arXiv190407224S}. 

The required material strength for CNEOS-2014-01-08 suggests a composition tougher than that of typical iron meteorites or else it would have not survived the ram pressure down to an altitude of 18.7 kilometers where its brightest flare was observed. The last flare occurred at a ram pressure of 194 MPa, implying that the yield strength was in the range of 113-194 MPa.

We computed the ram pressure of each of the 273 bolides in the CNEOS database with values for peak brightness altitude and speed. Interestingly, CNEOS 2014-01-08, with a ram pressure of 194 MPa at peak brightness, has the highest material strength of all 273 bolides. The second highest tensile strength is smaller by more than a factor of 2, namely 81 MPa for the 2017-12-15 13:14:37 bolide. The third highest tensile strength, 75 MPa, belongs to the 2017-03-09 04:16:37 bolide, which we identified as a possible interstellar meteor candidate \citep{2019arXiv190407224S}. 

Of course, this result does not imply that the first interstellar meteor was artificially made by a technological civilization and not natural in origin \citep{2021arXiv211015213L}. Iron meteorites make about a twentieth of all space rocks arriving on Earth. Their composition consists of 90-95\% iron with the remainder comprised of nickel alloys, including trace amounts of iridium, gallium and sometimes gold. In principle, an interstellar meteor could deliver exotic abundances of heavy elements, depending on the proximity of its birth place to a supernova or a merger event of two neutron stars. 

In conclusion, CNEOS 2014-01-08 is an outlier both in terms of its LSR speed (shared by less than 5\% of all stars) and its composition (tougher than all 272 other bolides in the CNEOS database).

%% file: tex/expedition.tex
\section{Plan for the Ocean Expedition}

\begin{figure}
  \centering
  \includegraphics[width=\linewidth]{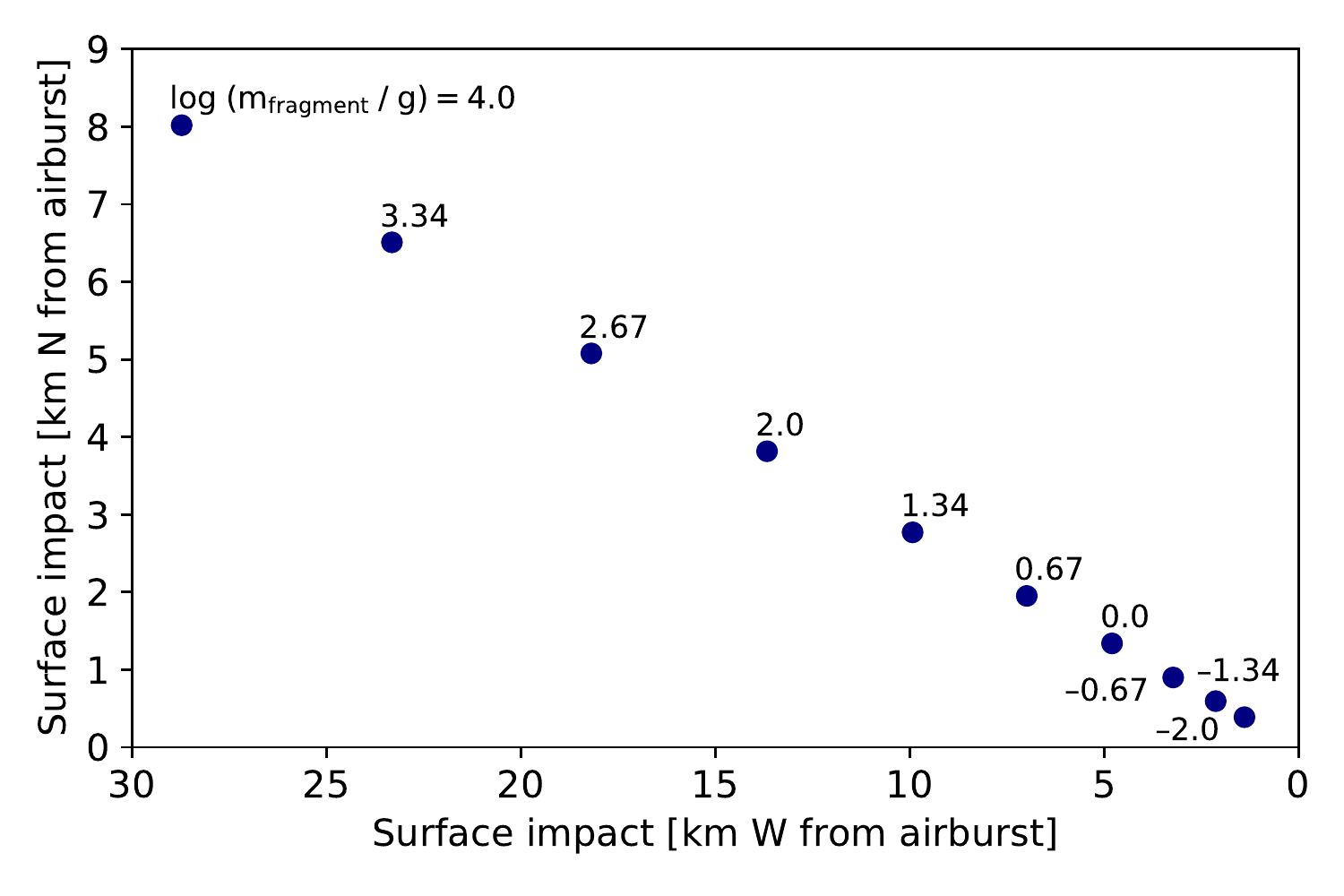}
    \caption{Predicted surface impact location of fragments as a function of mass relative to the airburst location, located at the origin (not accounting for wind).}
    \label{fig:surface_impact}
\end{figure}

Our plan is to mobilize a ship with a magnetic sled deployed using a long line winch. We will be operating approximately $\sim 300 \mathrm{\; km}$ north of Manus Island. The team will consist of seven sled operators, plus the scientific team. The goal of the expedition is to recover $\sim 0.1 \mathrm{\; mm}$ size fragments from the meteorite that exploded over the Bismarck sea in 2014. The recovered fragments will be carefully analyzed and will be shared with the global scientific community. We will tow a sled mounted with magnets, cameras and lights on the ocean floor inside of a $10 \mathrm{\; km} \times 10 \mathrm{\; km}$ search box. A number of sources have been used to narrow the search site to this relatively small search box. A sled, $\sim 2 \mathrm{\; m}$ long, $\sim 1 \mathrm{\; m}$ wide and $\sim 0.2 \mathrm{\; m}$ centimeters tall weighing about $\sim 55$ kg, will be towed along the seabed to sample for ferro-magnetic meteorite fragments from the CNEOS 2014-01-08.

The trip envisions ten survey days during which an the survey area will be sampled in down-current runs, corresponding to roughly westerly courses. The sled body will be a $\sim 10 \mathrm{\; cm}$ thick block of ultra high molecular weight polyethylene (UHMW) with $\sim 20 \mathrm{\; cm}$ tall steel runners on each side.  Rare earth magnets embedded in the UHMW are designed to capture any ferro-magnetic particles that the sled encounters. The tow speed will be about $\sim 3.3 \mathrm{\; km/hr}$.  On each 10-km run, an area 10 km by 1 meter will be disturbed. The sled will be recovered after each run to remove any particles attached to the magnets.  The maximum depth of deposition will be $\lesssim 5 \mathrm{\; cm}$ because we assess the seafloor in the region to be composed of calcareous ooze or clay and that the sedimentation rate is on the order of $1 \mathrm{\; mm} - 1 \mathrm{\; cm}$ per thousand years \citep{webb}. Each 11-km run will take $\sim 3$ hours. In seven survey days, considering turns and recovery times, we expect to have about 126 effective towing hours in the field or almost the equivalent of $\sim 40$ 10-km lines. This would disturb an area of $\sim 0.4 \mathrm{\; km}$ to a depth of no more than five cm. Another option being considered is towing a mesh net behind the sled to catch any larger particles. Part of the survey area will be a control sample in the region where we do not expect any fragments of CNEOS 2014-01-08, so we can properly distinguish the interstellar particles from the micrometeorite background.

Given the airburst explosion energy of $4.5 \times 10^{18} \; \mathrm{erg}$, the energy per unit area delivered to the seafloor $20.4 \mathrm{\; km}$ below the explosion was $f \cdot 8.6 \times 10^4 \mathrm{\; erg \; cm^{-2}}$, where $f$ is the transmission efficiency. Setting this value equal to the gravitational potential energy per unit area, $(\rho g h^2) / 2$, where $\rho \sim 3 \mathrm{\; g \; cm^{-3}}$ is the density of material on the ocean floor, we determine that the depth of ocean floor material displaced is $h = 7.6 \mathrm{\; cm} \sqrt{f}$. Depending on the sensitivity of available ocean-floor mapping instruments, we could potentially detect this change in depth over the blast wave radius of $\sim 20 \mathrm{\; km}$.

The best way to decipher anomalies is to gather additional data. We are currently planning an expedition to Papua New Guinea where we could retrieve the meteor’s fragments from the ocean floor. Studying these fragments in a laboratory would allow us to determine the isotope abundances in CNEOS-2014-01-08 and check whether they are different from those found in solar system meteors. Altogether, anomalous properties of interstellar objects like CNEOS-2014-01-08 and `Oumuamua, hold the potential for revising conventional wisdom on our cosmic neighborhood. The expedition to the ocean floor around Papua New Guinea will illustrate metaphorically how scientific evidence expands our island of knowledge into the ocean of ignorance that surrounds it. 